\documentclass[twocolumn,showpacs,aps]{revtex4}

\usepackage{graphicx}
\usepackage{bm}

\begin{document}
\title{Comment on ``Security proof for cryptographic protocols based only on monogamy of Bell's inequality violations"}
\author{Won-Young Hwang$^{1,2}$
and Oleg Gittsovich$^{2}$}
\affiliation{$^{1}$ Department of Physics Education, Chonnam National University, Gwangju 500-757, Republic of Korea\\
$^{2}$ Department of Physics and Astronomy, Institute for Quantum Computing, University of Waterloo, Waterloo, Ontario, Canada  N2L 3G1}
\begin{abstract}
Recently, Pawlowski [Phys. Rev. A {\bf 82}, 032313 (2010)] claimed to have proven the security of a quantum key distribution by using only the monogamy of Bell's inequality violations. In the proof, however, he tacitly assumed that the eavesdropper's outcome is binary. The assumption cannot be justified because Eve's (eavesdropper's) power can only be limited by natural principle. We provide a counter-example for a step of the proof.
\end{abstract}
\maketitle
In Ref. \cite{Paw10}, the author presented arguments to prove the security of a quantum key distribution by using only the monogamy of Bell's inequality violations. In the arguments, it is assumed that Eq. (11) implies Eq. (10), that is,
\begin{equation}
P_B > P_E \Rightarrow I(A,B) > I(A,E)
\label{A}
\end{equation}
Here, $P_E$ ($P_B$) is Eve's (Bob's) guessing probability for Alice's bit, and $I(A,B)$ ($I(A,E)$) is the mutual information between Alice and Bob (Alice and Eve).
If the outcome of each participant is binary, Eq. (\ref{A}) is valid. However, there is no reason why Eve's outcome has to be binary; Eve is assumed to have unlimited power within natural principle. There is no natural principle to limit number of the Eve's outcomes.

Without the assumption that Eve's outcome is binary, Eq. (\ref{A}) does not follow. There is an example in which Eq. (\ref{A}) is violated indeed. (The idea here is ``to know the small part clearly gives more information than to know the large part ambiguously does".)
\begin{eqnarray}
P_{000} &=& 1/4 -\epsilon \nonumber\\
P_{110} &=& \epsilon \nonumber\\
P_{001} &=& 1/8 \nonumber\\
P_{011} &=& 1/8 \nonumber\\
P_{102} &=& 1/8 \nonumber\\
P_{112} &=& 1/8 \nonumber\\
P_{003} &=& \epsilon \nonumber\\
P_{113} &=& 1/4 -\epsilon
\label{B}
\end{eqnarray}
Here, $P_{ijk}$ denotes a joint probability that Bob's outcome is $i$, Alice's outcome is $j$, and Eve's outcome is $k$, respectively. Alice and Bob's outcomes are binary, but Eve has four outcomes. All $P_{ijk}$'s not shown here are zero. $\epsilon$ is a small positive quantity, say $1/100$.
From Eq. (\ref{B}), we get
\begin{eqnarray}
P_B &=& 0.75  \hspace{1cm} P_E= 0.75 -2 \epsilon
\label{C}
\\
I(A,B) &\approx& 0.19  \hspace{1cm} I(A,E)= 0.50- \epsilon^{\prime},
\label{D}
\end{eqnarray}
where $\epsilon'= (1/2) H[4\epsilon] $ is also a small positive quantity. Here, $H$ is the binary entropy function. We can see that Eqs. (\ref{C}) and (\ref{D}) largely violate Eq. (\ref{A}).

To summarize, without the unjustifiable assumption that Eve's outcome is binary, Eq. (\ref{A}) is not valid. Therefore, security is not guaranteed by the arguments in Ref. \cite{Paw10}.

This study was supported by Basic Science Research Program through the National Research Foundation of Korea (NRF) funded by the Ministry of Education, Science and Technology (2010-0007208).

\end{document}